# Multiscale Co-Design Analysis of Energy, Latency, Area, and Accuracy of a ReRAM Analog Neural Training Accelerator

Matthew J. Marinella*, *Senior Member, IEEE*, Sapan Agarwal*, *Member, IEEE*, Alexander Hsia, Isaac Richter, *Member, IEEE*, Robin Jacobs-Gedrim, John Niroula, Steven J. Plimpton, Engin Ipek *Member, IEEE*, Conrad D. James *Member, IEEE*

*Abstract*—Neural networks are an increasingly attractive algorithm for natural language processing and pattern recognition. Deep networks with >50M parameters are made possible by modern GPU clusters operating at <50 pJ per op and more recently, production accelerators capable of <5pJ per operation at the board level. However, with the slowing of CMOS scaling, new paradigms will be required to achieve the next several orders of magnitude in performance per watt gains. Using an analog resistive memory (ReRAM) crossbar to perform key matrix operations in an accelerator is an attractive option. This work presents a detailed design using a state of the art 14/16 nm PDK for of an analog crossbar circuit block designed to process three key kernels required in training and inference of neural networks. A detailed circuit and device-level analysis of energy, latency, area, and accuracy are given and compared to relevant designs using standard digital ReRAM and SRAM operations. It is shown that the analog accelerator has a 270x energy and 540x latency advantage over a similar block utilizing only digital ReRAM and takes only 11 fJ per multiply and accumulate (MAC). Compared to an SRAM based accelerator, the energy is 430X better and latency is 34X better. Although training accuracy is degraded in the analog accelerator, several options to improve this are presented. The possible gains over a similar digital-only version of this accelerator block suggest that continued optimization of analog resistive memories is valuable. This detailed circuit and device analysis of a training accelerator may serve as a foundation for further architecture-level studies.

## I. Introduction

NEURAL networks have gained renewed, widespread attention in recent years. This is due in large part to the development of Deep Neural Networks (DNNs), which have demonstrated significantly better classification on image recognition and other datasets than previous techniques [1], [2]. Advances in hardware played a central role in enabling DNNs, which often have >10$^7$ parameters, to be trained in a reasonable time. Between the mid-1980s when backpropagation was introduced and the present, the power-normalized performance (e.g. GOPS/W) of computing hardware has increased by about six orders of magnitude [3]. In addition, the parallel nature of DNNs allows favorable mapping of neural networks to modern multi-core CPUs and GPUs.

Although CMOS continues to scale, frequency scaling ended around 2003 because voltage scaling slowed drastically and Dennard constant power density scaling ended [4]. At this point, single thread performance improvements dramatically slowed. CMOS voltages are presently reaching fundamental limits, and hence precluding future frequency scaling of dense transistors. Transistor dimensions continue to scale, but due to power density limits, voltage and frequency are dynamically controlled on-chip. Nevertheless, additional transistors have enabled some performance increases due to multiple cores, additional cache, and specialized blocks. Performance per watt gains will likely continue for about a decade as a result specialization and heterogeneous integration of memory.

Once the gains from CMOS scaling and heterogeneous integration have been exhausted, non-traditional techniques will be required to continue the gains in computing performance. In this work, we propose the use of an analog module which can efficiently perform a vector matrix multiply (VMM), matrix vector multiply (MVM), and an outer product update. These operations are typically bottlenecks in training of neural networks, and this module can improve their efficiency by several orders of magnitude when floating point precision is not required by an algorithm.

Using ReRAM as an analog programmable resistor module presents a significant design challenge: device properties can affect the algorithm level accuracy. The efficient separation of device, circuit, architecture, and algorithms enabled by the traditional VLSI methodology is no longer sufficient. In order to develop and analyze the analog neural training accelerator block, we have utilized a co-design methodology that is

Submitted for review on May 14, 2017. This work was funded by Sandia National Laboratories Hardware Acceleration of Adaptive Neural Algorithms (HAANA) Grand Challenge Laboratory Directed Research and Development (LDRD) Project. Sandia National Laboratories is a multimission laboratory managed and operated by National Technology and Engineering Solutions of Sandia, LLC., a wholly owned subsidiary of Honeywell International, Inc., for the U.S. Department of Energy's National Nuclear Security Administration under contract DE-NA0003525. M. J. Marinella, A. Hsia, R. Jacobs-Gedrim,

S.J. Plimpton, and C.D. James are with Sandia National Laboratories, Albuquerque, NM, 87185-1084. (email: mmarine@sandia.gov).

S. Agarwal is with Sandia National Laboratories, Livermore, CA 94550 (email: sagarwa@sandia.gov)

E. Ipek and I. Richter are with the Dept. of Electrical and Computer Engineering, University of Rochester, Rochester, NY, 14627. (email: engin.ipek@rochester.edu, isaac.richter@rochester.edu)

*M. J. Marinella and S. Agarwal contributed equally to this work

developed to enable the use of measured device data to predict accuracy at the algorithm level. The co-design software, CrossSim, is open source and available online[5]. Key contributions of this paper are:

1. The design of an analog ReRAM accelerator block including analog and digital components using a state of the art 14/16nm-node commercial process development kit (PDK) for the processing of three key kernels required for neural network training: vector matrix multiply, matrix-vector multiply, and outer product update.
2. The design of alternative digital-only SRAM and ReRAM accelerator components for comparison.
3. Comparative energy, latency, and area analysis of the three versions: i) analog-ReRAM, ii) digital-ReRAM, and iii) SRAM (CMOS-only) for each of three kernels.
4. The extensive analysis of analog accelerator training accuracy based on experimental properties of ReRAM.

The remainder of this paper is organized as follows. Section II provides a brief background on devices, algorithms, and related work. Section III describes the analog neural training accelerator circuit block architecture, followed by the energy, latency and area analysis in Section IV. Sections V and VI details the measurement and evaluation of ReRAM for the accelerator. Section VII discusses future challenges.

## II. BACKGROUND

### A. Multilayer Perceptron Network

Neural networks can process pattern recognition tasks such as image recognition and natural language processing more accurately than traditional machine learning techniques. The multilayer perceptron network (MLP) algorithm is a common element in DNNs, and is used to assess training in this work. The proposed accelerator can map to a number of neural network algorithms. The basic element of a neural network is the neuron, which outputs the weighted sum of inputs put through an activation function (typically a sigmoid). The number of neurons and weights depend on the structure of the data being analyzed. For example, the MNIST dataset is composed of black and white images of handwritten digits 0-9 with 784 pixels each [6]. The task of the network is to recognize the image as a digit. Hence the input layer of the network must contain 784 inputs and the output layer must have 10 outputs. The size and number of intermediate layers can be used as a parameter to optimize the network accuracy.

Before the neural network can recognize patterns (inference), all of the weights are trained by cycling through each element of a training dataset, and adjusting weights depending on the error as defined by the training algorithm. This work uses the backpropagation of errors for training, which calculates and error attributed to each weight, starting with the output layer.

### B. CMOS-Based Neural Network Accelerator Work

Neural network training and inference are computationally intensive, which has spurred interest in acceleration of both using specialized hardware. Google has recently provided a performance analysis of their tensor processing unit (TPU) being used for inference with deep MLPs, convolutional neural nets (CNNs), and long short-term memory deep networks with as many as $10^8$ weights [7]. Google's accelerator achieves a 30x improvement in performance per watt over the contemporary GPU (Nvidia K80), with an estimated gain of 70x if the memory system was upgraded to that of the GPU. At the die level, the TPU can achieve about 2.3 TOPs/watt (or about 430 fJ/op) with 8 bit fixed precision. This likely represents the most practical application of a specialized DNN accelerator, saving the construction of several Google datacenters.

DaDianNao represents a set of accelerators which have been designed for DNN and CNN inference [8]. They have been architected to make memory movement as efficient as possible. A cycle-accurate simulation of the DaDianNao version estimates of gives an estimated 600 GOPS/W (3 pJ/op) processing a deep MLP with 16 bit weights. The chip has not been fabricated, but the design was completed through layout, so performance estimates should be reasonably accurate.

A key conclusion from the study of state of the art neural CMOS-based accelerators is that 8- or 16-bit operations are on the order of 1 pJ at the chip-level. Hence, for an analog accelerator that relies on new device technologies to be viable, an improvement over the state of the art of at least 10x is needed. Since we expect the analog accelerator to take about 5-10 years to develop, it should be assumed that it will need to achieve an order of magnitude over state of the art in that timeframe. The state of the art is rapidly advancing, and it is reasonable to expect with CMOS and integration of emerging memories, that within the next 5-10 year period this energy per operation will improve an order of magnitude. *Hence, the target for an analog training accelerator is 10 fJ per operation, or 100 TOPs/W*. The operations which must meet this threshold are multiply, accumulate, and update a weight matrix. Therefore, this target can be expressed as 20 fJ per MAC, or 50 TMAC/W to be consistent with metrics for analog blocks in the literature.

### C. Analog Neural Accelerator Related Work

Several architectural studies of neural accelerators have appeared recently. The ISAAC architecture is a full neural execution unit similar to DaDianNao but using ReRAM crossbars to store and process weights for CNN inference [9]. PRIME is a new pipelined architecture and a method of efficiently processing neural network inference with analog weights. PRIME provides an energy advantage of three orders of magnitude [10].

RENO is another neural accelerator architecture utilizing digital and analog-ReRAM crossbar operations to perform inference on neural networks [11]. Up to a 177x performance gain and 185x energy savings are gained compared to a CPU core. Hasan *et al* compares a RISC-based processor with a digital CMOS, and analog ReRAM accelerator for image recognition and edge detection tasks using 40nm CMOS technology parameters [12]. The CMOS/SRAM accelerator gains about 3 order magnitude in power efficiency and up to five orders of magnitude for the analog ReRAM version.

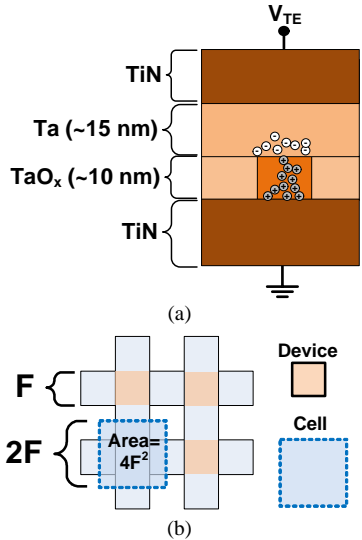

Fig. 1. Sandia TiN/Ta/TaOx/TiN ReRAM device.

In each case, the results depend strongly on the algorithm. Performance per watt and latency were not reported so absolute energy per operation cannot be compared to CMOS results above (on the order of 1pJ per 8-bit precision operation).

Numerous experimental demonstrations of analog vector matrix multiply have been reported recently. For example, Chakraborty *et al* recently experimentally demonstrated a 500 nm CMOS with monolithically integrated ReRAM crossbars that can perform a vector matrix multiply of a 24 x 36 matrix [13]. This promising work shows feasibility of analog crossbar VMM. A drawback is that analog ReRAM currents were high (for example, in the 10 µA range so a large-array parallel operation needed to compete with CMOS is still prohibitive).

From the preceding discussion, there is still a wide range of reported gains in performance and energy per operation. More precise circuit and device analysis are needed, which should utilize a detailed co-design philosophy. To the best of our knowledge the analysis of energy, latency, area, and accuracy for a training and inference accelerator using a commercial 14/16nm PDK and experimental $TaO_x$ device data is unique.

### D. Resistive Memories

A number of two-terminal resistance change memories are currently being explored for next generation high density, high endurance storage class memories (SCM) [14] and embedded memories. Chief among these are redox-based random access memory (ReRAM), phase change (PCRAM), conducting bridge memory (CBRAM), and Ferroelectric Tunnel Junction (FTJ) [15]. In addition, more novel nonvolatile devices, such as those based on Lithium-Ion battery physics have been also demonstrated [16]. In this work, we use metal oxide-based ReRAM, also often referred to as Ox-RAM as an example device. However, the analog neural training accelerator can use any resistance change devices that meet the voltage, current, and variability specifications discussed below.

Our assessment of analog ReRAM properties is largely based on Sandia's TiN/Ta/TaOx/TiN ReRAM cell which is shown in Fig. 1(a). Process details for these devices are discussed previously [17], [18]. The device operation is preceded by an electroforming step, which serves to create a small, high conductivity region and further anneal the switching film. Electroforming typically occurs at $V_{TE}$=+2 to 5V using either a voltage ramp or a voltage pulse train, ending when a maximum current is reached. After electroforming, the device is reset by applying a negative voltage $V_{TE} \approx$ -1 to -2.5V, with a pulse length ranging from 10ns to 1 µs.

Similar Ta/TaOx-based cells described in the literature have demonstrated endurance as high as $10^{12}$ cycles with estimated retention of 10 years [15], [19]. Reliable operation with write currents in the range of 50 nA and high resistance state (HRS) down to 1 pA has been demonstrated with precise barrier engineering [20]. Oxide ReRAM as small as 10 nm have been demonstrated [21]. With a $4F^2$ cell layout (Fig. 1(b)) and monolithic layering it will be possible to achieve densities on the order of 10-100 Tbit/cm$^2$.

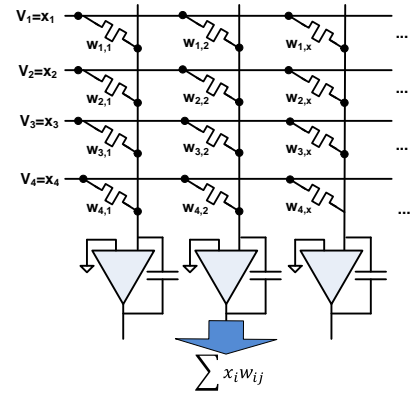

Fig. 2. Analog vector matrix multiply using a resistive memory crossbar. Each column performs the mathematical operation of multiply of the weight and input voltage using Kirchoff's voltage law. The currents along the column sums are summed using Kirchoff's current law, resulting in the sum.

### III. ACCELERATOR ARCHITECTURE

Three key computational kernels underlie many different neural algorithms including backpropagation, sparse coding, and restricted Boltzmann machines. The kernels are:
1) the parallel read, or vector matrix multiply (VMM),
2) transpose parallel read, or matrix vector multiply (MVM),
3) the parallel write or rank one outer product update.

Each of these can be performed with an analog ReRAM crossbar. Fig. 2 illustrates this concept of a crossbar vector matrix multiply. Kirchoff's voltage law provides the product, $x_iw_{ij}$ and the current law provides the sum $\sum x_iw_{ij}$. Each vector element $x_i$ is represented by a voltage and weight $w_{ij}$ by a conductance. This entire operation can be done in parallel, as opposed to a traditional system which must multiply each element serially and accumulate the answer. The transpose matrix vector multiply can also be done with the crossbar by driving columns and measuring the rows. Controlling both rows and columns and using time and voltage encoding also allows us to update each weight in the crossbar (rank-1 update) as a single parallel operation, as discussed below. These kernels form the foundation of a neural accelerator [22].

Performing these operations in parallel with an NxN crossbar reduces the total operations from $O(N^2)$ to $O(N)$ inputs or outputs. Hence, although ReRAM elements may not be individually as fast or energy efficient as digital SRAM cell, these parallel operations can be performed faster on a crossbar than digital system. When integrated as a hybrid analog digital system, the tradeoff between energy efficiency and system flexibility can be optimized. A crossbar neural core performs matrix operations, a digital core processes the results, and cores are connected through a routing network [13, 14].

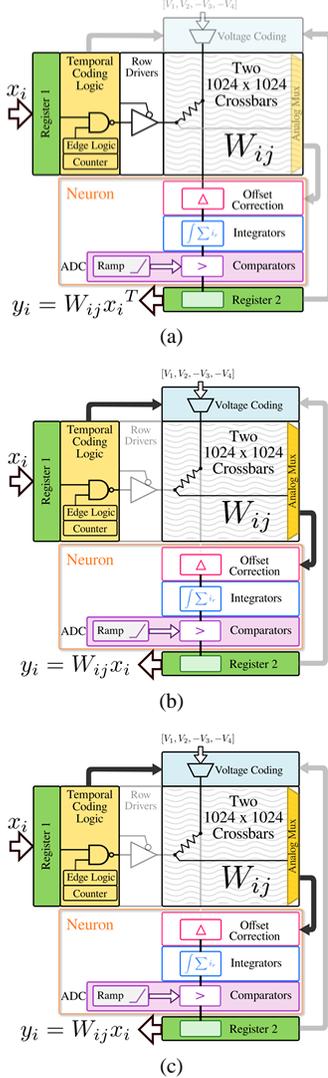

Fig. 3. The design of the neural core is highlighted showing the three key computational kernels: (a) vector matrix multiplication, (b) matrix-vector multiplication, and (c) outer-product update. Gray sections represent circuitry not in use in a particular operation.

We now focus on the details of the analog neural core. In the following, we first explain the three key operations of the neural core illustrated in Figs. 3(a), (b), and (c) respectively. Then we explain more specific details of the electronic read and writing of analog weights.

*A. Vector Matrix Multiply*

First, consider the vector matrix multiply illustrated in Fig. 3(a). In this operation, a vector input, $x_i$ is loaded in the digital register 1 from the bus. In the crossbar VMM, the rows are driven with the signal representing the vector and the results of the multiply accumulate are read on the columns. In particular, the vector $x_i$ is encoded into variable length pulses that are applied to the crossbar using the coding logic described below. Row drivers provide a constant voltage (~0.8V) pulse of variable length. An additional "offset correction" row is added to the crossbar and total currents are integrated as the analog sum of charge through the column in the "integrator" block. Finally, the analog voltage output is sent through the analog to digital converter (ADC), the design of which is detailed below. The result is stored in register 2.

*B. Matrix Vector Multiply*

The neural core can perform a matrix vector multiply as illustrated in Fig. 3(b). This operation is similar to the VMM but requires *driving the columns* and *reading the rows*. The input vector is loaded into register 1 and converted into a pulsed signal. In this case the temporal signal is routed to the columns, which are driven with a constant voltage using the "voltage coding" block above the array. It should be noted that in this case, the "voltage coding" block is being used only to provide a constant variable length voltage pulse. The "analog mux" allows us to integrate the current from the rows while reusing the neuron circuitry (offset correction, integrators, and comparators) used in the VMM. The final digital output is stored in register 2 before sending to the external digital core.

*C. Outer Product Update*

The last key operation performed by the neural core is a parallel outer product update (Fig. 3(c)). The neural network weight set $W_{ij}$ represented by the conductances is updated by increments defined by values $x_i \otimes s_j$. In order to accomplish this, the vector $x_i$ is input into register 1 and converted to a temporal signal with the "temporal coding logic" block. The vector $s_j$ is input into register-2 and coded in voltage using the "voltage coding block". This hybrid voltage-temporal coding avoids increasing the update time as $2^{(2\times bits)}$ ns if only time coding was used. Pulse lengths and voltages must be carefully chosen to compensate for writes that depend nonlinearly on the voltage or pulse length. The final result is the update of the weight set such that $W_{ij\text{-updated}} = W_{ij} + x_i \otimes s_j$.

*D. Serial Reads and Writes*

In order to initialize or copy an array, serial reads and writes are needed to access each resistive memory individually. The parallel hardware described above can be used for serial operations by driving only a single row at a time. If needed, longer read pulses or a smaller capacitor can be used with the integrator for serial reads to improve the dynamic range.

*E. Encoding and Reading Analog Weights as Conductances*

*1) Negative Weights*

To represent both positive and negative matrix values with a resistive device, the difference between two memory elements is taken as illustrated in Fig 4. When a positive read pulse is applied to a positive weight, the opposite negative pulse is applied to the corresponding negative weight. This ensures that the total current at the integrator will be the difference between the two. The negative weights are initialized to a fixed reference resistance at the midpoint of the conductance range so they subtract a fixed offset.

Subtracting reference weights in analog is an expensive operation as it roughly doubles the current used. However, this maximizes the dynamic range of the integrator by reducing the amount of charge that needs to be integrated. This reduces the required capacity of the integrator capacitor, maximizing the amount of voltage output for a given amount of charge. The ADC and integrator still dominate the energy and so doubling the read current is does not dominate the overall energy.

Any variations in the reference resistors will translate into shifts in the zero point of each weight. This can be compensated for by an initial calibration of the weights, or can be ignored and considered part of the random initialization of the weights. We deliberately chose to use a full array of reference weights, rather than more compact schemes that only use a single column and an op-amp as in [23]. This is to ensure that an identical reference weight is used for both VMM and MVM, making the system more robust to variability in the reference weights. It also eliminates issues of variability in the required op-amp. As seen later, the area cost is reasonable as the driver circuitry is shared and the extra array fits over the required drivers.

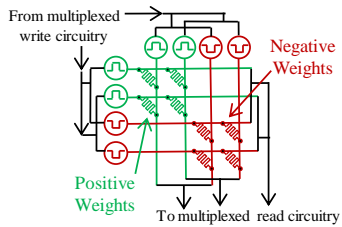

Fig. 4. Negative weight representation scheme.

*2) Temporal Coding Drivers*

Digital inputs are encoded into variable length pulses by ANDing each bit of the input register with a pulse of the appropriate length as illustrated in Fig. 5. The pulses are generated once using a counter for all of the drivers in a given array. The rows are then driven by positive or negative voltages based on the sign of the input. It is also possible to disconnect the driver or give a high-Z input when other crossbar operations are running. The drive circuitry (and register 1) is synthesized from Verilog and outputs three control signals that connect the row (or column) to one of three possible voltages: a positive voltage, a negative voltage, or a standby/ground voltage as illustrated in Fig 6(a). The particular voltage connected to the row (or column) is sourced by a pass transistor connected to the appropriate voltage rail. The precise voltage values depend on whether the driver is used for a read or a write and are selected by connecting the rail to the appropriate voltage source.

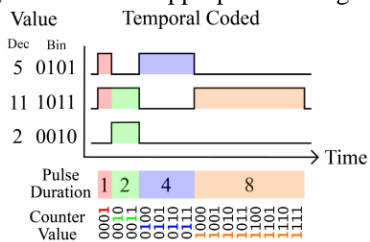

Fig. 5. Input pulses are encoded by ANDing the binary values with pulses of the appropriate length.

The analog array requires higher drive voltages (~1.8V) than the digital logic CMOS nominal voltage of 0.8V. In order to use low voltage control logic to activate the high voltage pass transistors connecting the rails, a level shifter is needed to step up the voltage as illustrated in Fig 6(b). Complementary outputs from the level shifter drive both the positive and negative arrays (see Fig 6(c)).

*3) Voltage Coding for Parallel Weight Update*

The voltage coding driver encodes drive signals for each column's PMOS and NMOS devices by connecting the desired input voltage to the column using the same design as the read driver, Fig. 6(c), but with one voltage rail for each level. The control circuitry and register 2 (Fig. 7) is synthesized in Verilog. It should be noted that because row and column inputs can be either positive or negative, a single write phase is insufficient. If the row voltage is positive, only a negative column voltage will cause a resistive memory to write. Four write phases are needed to capture all four possible combinations or row and column voltages (++, +-, -+, --). This halves the size of the per column voltage driver as positive and negative voltages are done in separate phases. The write pulses should be structured so that unselected devices see at most 1/3 of the write voltage following a V/3 write scheme. For instance, consider the ++ phase. Row and col. drivers that are fully ON have $+V_{write}/2$, and $-V_{write}/2$ respectively while row and column drivers that are OFF have $-V_{write}/6$ and $+V_{write}/6$ respectively, giving a maximum unselected voltage of $+/-V_{write}/3$.

*4) "Neuron" Circuitry*

After applying the input pulses, the outputs are integrated using a current conveyor based integrator and then digitized using a ramp based ADC [24] as illustrated in Fig 8. Current conveyors have large bandwidth, a virtual ground-like node, and low input impedance – which are desirable traits for an integrator. To save energy and area, all of the comparators share the same ramp generator and master counter. When a comparator triggers, it causes the counter value to be latched into its corresponding digital output buffer. Since it has to continually compare against the incoming ramp, the comparator cannot be of the common regenerative latch type and must also operate in continuous time. It needs a large transconductance to garner both the speed and the gain necessary to amplify the < 1 LSB voltage difference (~4mV) to the full 1.8V rails in about a 1ns. The current design uses the higher mobility NMOS for greater transconductance and relies on partial positive feedback (M3/M4) to boost the gain high enough in a single stage to generate a full rail swing from an input 1 LSB difference.

We deliberately do not include offset correction in the integrator or comparator. Instead, an extra offset correction row is added to the crossbar. This row is always active during a read and adds (or subtracts) a fixed amount of current from the integrator. By programming this row during an initial calibration step based on measuring zero input current, it can exactly subtract off the offsets due to the integrator and op-amp.

A key challenge is designing an integrator that can respond sufficiently to the high speed (~1ns) time dependent inputs while maintaining energy and area efficiency. This is mitigated by two design choices. First, a current conveyor enables faster response than a traditional integrator. Offset from the current conveyor is corrected using a bias weight, or digital offset. Second, a large load capacitance added to the columns (Fig. 8(a)) stores the initial current and limits the column voltage change while the integrator responds (the precise value of the capacitance is discussed below). This reduces the integrator response requirement.

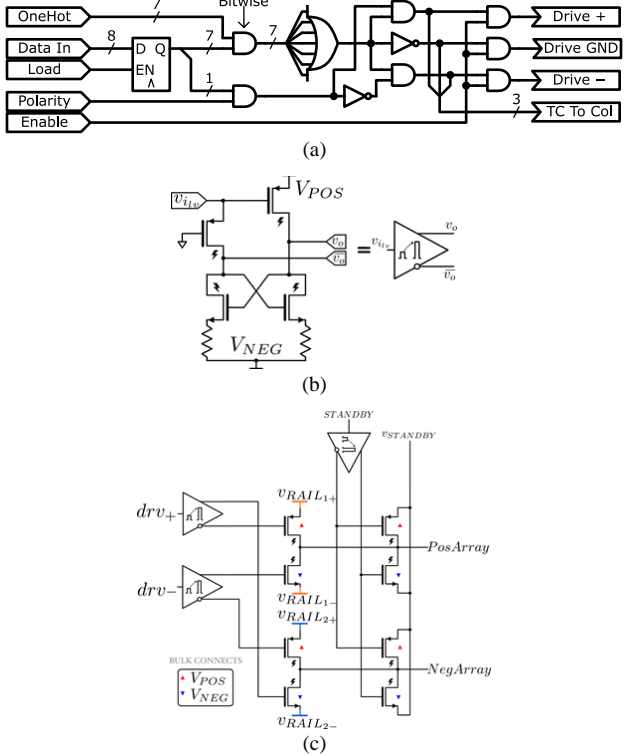

Fig. 6. (a) Digital logic for 8-bit temporal coding driver, including data buffer. "Data In" and "Load" place data into the buffer. One bit of data is for the sign. While driving, "OneHot" indicates the leading 1 of the counter indicating which bit is currently being encoded. "Polarity" allows the reversal of the positive and negative outputs, which is required during writes. When "Enable" is off, the drive outputs are off, causing the analog drivers to go into high-impedance. The "TC To Col" output sends temporal-coded signals to the voltage driver for use during column-driven reads. (b) Schematic of a level shifter which converts the logic-level (0 to $V_{DD}$) inputs to high-voltage outputs (±900 mV) sufficient for the gates of the high-voltage transistors driving the arrays by using positive feedback to increase the output voltages [25, 26]. The series resistances effectively increase the resistance of the NMOS transistors, causing a larger NMOS/PMOS mismatch and allowing the feedback to occur quickly with minimum-size devices. (c) Schematic of the circuitry used to drive both the positive and negative weight arrays.

## IV. Circuit Block Efficiency

In this section, we analyze the area, energy, and delay of the analog neural core and compare it to accelerator core designs using digital ReRAM and digital CMOS-only (SRAM). In order to model the architecture, we use a commercial 14/16nm FinFET PDK for digital and analog transistor properties. All logic operates on a 1 GHz clock. Key properties used in the model are summarized in Table I (with approximate values reported for proprietary numbers). We consider an accelerator built around a 1024x1024 array with 8-bit inputs and outputs with one being a sign bit. For digital comparisons, we use eight bit weights. For the outer product update, we limit the bit precision to 8 bits x 4 bits. Ref. [27] shows that as low as 2 bits x 2 bits can be used to achieve ideal numerical accuracy. For the updates, we assume a worst-case energy where all memory elements must be updated.

In Tables II-IV, we consider two additional accelerator architectures with 4-bit and 2-bit inputs and outputs. For the 4-bit version, the outer-product update is 4 bits x 2 bits. The 2-bit version is effectively one data bit and one sign bit and has a 2 bit x 2 bit outer product update. The length of the read pulse and write pulses are increased to 7 ns in the 2-bit architecture to ensure that there is sufficient charge integrated during a read and that sufficiently strong writes can be performed on the resistive memories. In all cases, the weights must remain at 8 bits to accumulate information over many training cycles.

The ReRAM is assumed to have an on-state resistance of 100MΩ [20]. This high resistance is critical for enabling parallel operation, as the maximum current in a wire needs to be limited to less than ~10 µA to avoid unacceptable line voltage drops (>20mV) [28] and stay within the current drive capacity of minimum sized transistors. We also place every ReRAM in series with an access device to prevent current flow at low voltages and enable parallel writes [29]. The access device is assumed to be a symmetrized diode following [29]:

$$I = sign(V) \times I_O \times \left(\exp\left(\frac{|V|}{V_O}\right) - 1\right) \quad [1]$$

$I_o$ is 8.7x10$^{-18}$A and and $V_o$ is 0.037V.

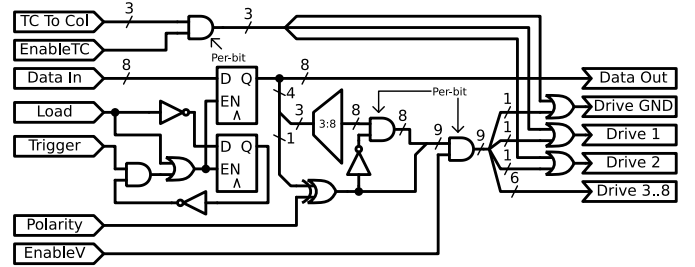

Fig. 7. Schematic of the voltage coding drivers. "Data In" and "Load" place data into the buffer. During a read, "Data In" has the output of the counter indicating the current state of the ADC ramp. When the ADC comparator signals equality, it asserts "Trigger", which causes "Data In" to be latched, and also causes the trigger register to trip. The trigger register is cleared during the next "Load". During column-driven reads, the "TC To Col" input contains the temporal-coded driver states, and is directed to the drive outputs when "EnableTC" is asserted. During writes, "EnableV" is asserted, which causes the relevant rail to be driven, if "Polarity" matches the stored polarity. Although the buffer can store 8 bits, only 1 sign bit and 3 decoded bits are used when performing writes. All 8 bits are used during reads as possible ADC outputs.

For a digital comparison, we assume that each weight is 8 bits, requiring a 1024x1024x8 = 1 MB of storage for each array. To balance latency vs area, we consider 256 multiply accumulators (MAC) in parallel and use one 1024x8 bit register to store the input data and use the MAC registers to store the output data.

The area, energy, and delay of all components are described below and the results are summarized in Tables II-IV. In analog ReRAM, the read (VMM) and read transpose operations (MVM) require the same amount of time and energy and the multiplication and accumulate (MAC) operations are free. In the digital CMOS-only version, the read and read transpose are different due to the memory array architecture. As part of a digital outer-product update, the array must be read, the outer product calculated and added to the weight, and then the weight updated, incurring the cost of read, write and MAC operations.

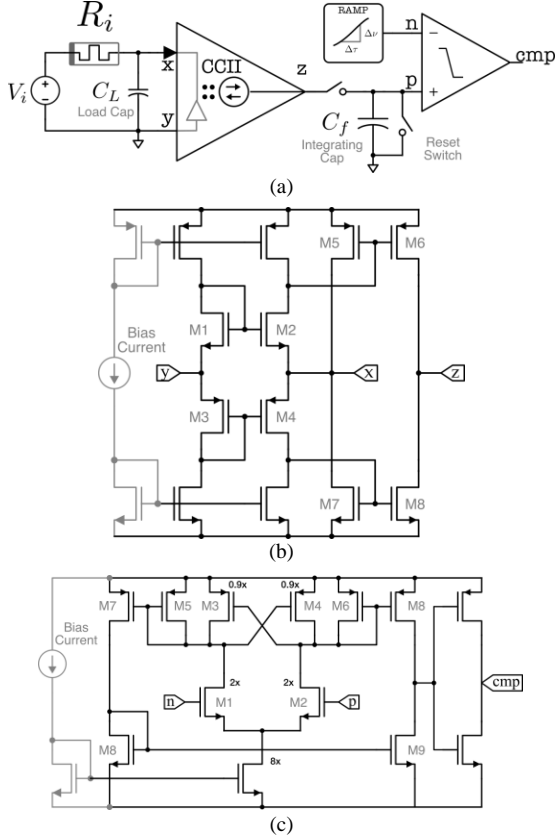

Fig. 8. (a) Schematics of the "neuron" circuitry and (b) current conveyor based integrator. There is a virtual ground between the x and y nodes set by the translinear loop of M1-M4 and a low impedance of ~1/g_m at node x. Since there is no global feedback, the conveyor is not bound by the same gain-bandwidth tradeoffs seen in traditional capacitive feedback configurations: greater bandwidth for integrating fast pulses can be obtained for smaller currents. (c) A schematic of the comparator. The leftmost grey bias current transistors are shared across multiple current conveyors or integrators.

| Quantity | | Value |
|---|---|---|
| Interconnect | Full Pitch($W_{M1\_Pitch}$) | 64 nm |
| | Capacitance | ~200 aF/μm |
| | Resistance | ~30 Ω/μm |
| Logic Transistor | Area | ~0.04 μm² |
| | Voltage | 0.8 v |
| High-Voltage Transistor | Area | ~0.35 μm² |
| | Voltage | 1.8 v |
| Crossbar | Dimensions ($n_{rows} \times n_{cols}$) | 1024 × 1024 |
| | Minimum Pulse Width | 1 ns |
| ReRAM & Select Device | ReRAM ON/OFF Ratio | 10 |
| | Capacitance ($C_{ReRAM}$) | 35 aF |
| Analog ReRAM & Select Device | On State Read Current | 1 nA ($R_{on}$ = 100 MΩ) |
| | On State Write Current | 10.3 nA ($R_{on}$ = 100 MΩ) |
| | Read Voltage | 0.785 V |
| | Write Voltage | 1.8 V |
| Binary ReRAM & Select Device | On State Read Current | 98 nA ($R_{on}$ = 1 MΩ) |
| | On State Write Current | 846 nA ($R_{on}$ = 1 MΩ) |
| | Read Voltage | 0.954 V |
| | Write Voltage | 1.8 V |
| Digital Array | Weight Precision | 8 bits |

Table I: Model properties and assumptions.

In order to estimate the area of different components, we count the number of transistors and multiply by an average area per transistor. High voltage transistors have a 2.6X higher gate pitch, 2X as many fins, and need 4X as much buffer space, resulting in an 8X larger area.

| Component | Area 8 Bit | Area 4 Bit | Area 2 Bit |
|---|---|---|---|
| Analog | | | |
| Arrays | 8,600 μm² | 8,600 μm² | 8,600 μm² |
| Temporal Driver Analog Transistors | 7,180 μm² | 7,180 μm² | 7,180 μm² |
| Temporal Driver Cache and Control Circuitry | 8,900 μm² | 5,100 μm² | 3,100 μm² |
| Voltage Drivers Analog Transistors | 26,000 μm² | 8,600 μm² | 8,600 μm² |
| Voltage Drivers: Cache and Control Circuitry | 18,000 μm² | 10,000 μm² | 7,100 μm² |
| Integrators | 6,600 μm² | 6,600 μm² | 6,600 μm² |
| ADCs | 5,850 μm² | 5,850 μm² | 5,850 μm² |
| Analog Routing | 2,900 μm² | 2,900 μm² | 2,900 μm² |
| Digital | | | |
| Array: 1MB ReRAM | 76,000 um² | 76,000 um² | 76,000 um² |
| Array: 1MB SRAM | 775,000 μm² | 775,000 μm² | 775,000 μm² |
| Multiply & Accumulate (256 in parallel) | 54,000 μm² | 35,000 μm² | 23,000 μm² |
| Input Buffers | 7000 μm² | 3500 μm² | 1750 μm² |
| Totals | | | |
| Analog ReRAM Total | 75,000 μm² | 46,000 μm² | 41,000 μm² |
| Digital ReRAM Total | 137,000 μm² | 114,000 μm² | 101,000 μm² |
| Digital SRAM Total | 836,000 μm² | 814,000 μm² | 800,000 μm² |

Table II: Area breakdown.

### A. Analog Array

The area of the two arrays is given by the eqn:

$$A_{array} = 2 \times n_{rows} \times n_{cols} \times (W_{M1\_Pitch})^2 \quad [2]$$

where $W_{M1\_Pitch}$ is the M1 full pitch.

The 90% rise time for the array is $2.2 \times \tau_{RC}$, where $\tau_{RC}$ represents the time constant for a row, which is ~0.2 ns. This in negligible compared to the temporal driver delays.

The read energy of the array consists of the dynamic $CV^2$ energy and the static $IV$ energy. The energies are doubled to account for positive and negative weights arrays. In the temporal code, the lines can switch once per input bit minus the sign bit ($n_{bits,T} - 1$), and will switch 50% of the time on average. Assuming the inputs are randomly distributed, there is a 50% chance any bit is on and driving static current. Thus the total energy is:

$$E_{READ} = \frac{1}{2} \times 2 \times (n_{bits,T} - 1) \times n_{rows} \times C_{line} \times V_{READ}^2$$
$$+ \frac{2}{2} \times n_{rows} \times n_{cols} \times I_{READ} \times V_{READ} \times 1\text{ns} \times (2^{n_{bits,T}-1} - 1) \quad [3]$$

where $C_{line} = n_{cols} \times (C_{wire} + C_{ReRAM})$ and $C_{wire}$ is the capacitance per cell of the wire and $C_{ReRAM}$ is the combined capacitance of the ReRAM and access device.

The write cycle is divided into 4 phases, with one quarter of the devices being written in each phase. The devices that are written will see the full write voltage $V_{WRITE}$ and pass a write current $I_{WRITE}$, assuming the voltage drivers can hold the max write voltage. The unselected devices will see up to $1/3 V_{WRITE}$ and pass a negligibile amount of current as the applied voltage is below the select device threshold. Only one array is written and the reference array is left unchanged.

First, we consider the $CV^2$ energy. Across 4 cycles, there is a possibility of writing in a single cycle. During that write cycle, we assume the temporal driver has a 50% chance of being a 1 during any given bit. Thus, the setup energy at the start cycle is given by:

$$n_{rows} \times C_{line} \times \left( 3 \left( \frac{V_{WRITE}}{3} \right)^2 + \frac{1}{2} V_{write}^2 + \frac{1}{2} \left( \frac{V_{WRITE}}{3} \right)^2 \right) \quad [4a]$$

In two of the write phases the temporal drive will have a 50% probability of transitioning during the $(n_{bits,T} - 2)$ edges between each bit. Half will switch against $\pm \frac{V}{6}$ costing $C \left( \frac{V}{3} \right)^2$ and the other half will switch against $\pm \frac{V}{2}$ costing $\frac{1}{2} C \left( V^2 - \left( \frac{V}{3} \right)^2 \right)$ on average. Thus the transition energy is:

$$\frac{2}{2} \times n_{rows} \times (n_{bits,T} - 2) \times C_{line} \times \left( \frac{1}{2} \left( \frac{V_{WRITE}}{3} \right)^2 + \frac{1}{2} \times \frac{4}{9} V_{write}^2 \right) \quad [4b]$$

Finally, the I-V energy is:

$$\frac{1}{2} \times n_{cols} \times n_{rows} \times I_{WRITE} \times V_{WRITE} \times 1\text{ns} \times (2^{n\ bits,T-1} - 1) \quad [4c]$$

Thus the total write energy is the sum of 4(a-c).

| Component | Delay 8 Bit | Delay 4 Bit | Delay 2 Bit |
|---|---|---|---|
| Analog | | | |
| Array | 0.2 ns | 0.2 ns | 0.2 ns |
| Read: Temporal Driver | 128 ns | 8 ns | 8 ns |
| Read: ADC | 256 ns | 16 ns | 3 ns |
| Write: Temporal Driver×4 | 512 ns | 32 ns | 32 ns |
| Digital | | | |
| Read: 1MB SRAM | 4 µs | 4 µs | 4 µs |
| Read Transpose: 1MB SRAM | 32 µs | 32 µs | 32 µs |
| Write: 1MB SRAM | 4 µs | 4 µs | 4 µs |
| Read: 1MB ReRAM | 176 µs | 176 µs | 176 µs |
| Read Transpose: 1MB ReRAM | 176 µs | 176 µs | 176 µs |
| Write: 1MB ReRAM | 164 µs | 164 µs | 164 µs |
| Multiply and Accumulate (256 in parallel) | 4 µs | 4 µs | 4 µs |
| Totals | | | |
| Analog ReRAM Total | 1.280 µs | 0.080 µs | 0.054 µs |
| Digital ReRAM Total | 692 µs | 692 µs | 692 µs |
| Digital SRAM Total | 44 µs | 44 µs | 44 µs |

Table III: Latency Per Component. The total time is for a three step cycle, a VMM, a MVM, and an outer product update.

### B. Temporal Drivers

For each row, the temporal drivers consist of digital buffers, logic and analog drivers. The digital logic was designed in Verilog and then synthesized using standard cells to give an area of 8.6 µm² per row for 8-bit values. It includes data storage, register 1 in Fig. 3, which was synthesized as part of the control logic. The control logic operates with the following steps:
1) find the leading 1 from counter,
2) AND the result of 1) with the registers
3) OR the result of 2) to determine if the line should be driven
4) determine which sign is driven based on the stored sign bit and requested polarity, and
5) send the outputs to the voltage shifters.

The analog drivers illustrated in Fig. 6 require 20 high-voltage transistors, including both voltage shifters, which convert the logic-level signals to high-voltage for the drive transistors, as well as the drive transistors themselves requiring an area of 7 µm². The total driver area is multiplied by $\max(n_{rows}, n_{cols})$.

| Component | 8 Bit Energy | 4 Bit Energy | 2 Bit Energy |
|---|---|---|---|
| Analog | | | |
| Read: Array | 0.36 nJ | 0.13 nJ | 0.07 nJ |
| Write: Array | 1.66 nJ | 0.31 nJ | 0.22 nJ |
| Temporal Driver Analog Transistors (1 cycle) | 0.16 nJ | 0.08 nJ | 0.04 nJ |
| Temporal Driver Digital Logic (1 cycle) | 0.04 nJ | 0.02 nJ | <0.01 nJ |
| Voltage Driver Analog Transistors (4 cycle write) | 0.08 nJ | 0.08 nJ | 0.08 nJ |
| Voltage Driver Digital Logic (4 cycle write) | 0.02 nJ | 0.01 nJ | 0.01 nJ |
| Read: Integrator | 2.81 nJ | 0.15 nJ | 0.15 nJ |
| Read: ADC | 9.4 nJ | 0.59 nJ | 0.15 nJ |
| Analog Cross Core Communication | 0.08 nJ | 0.06 nJ | 0.06 nJ |
| Digital | | | |
| Read: 64 128kb SRAMs | 286 nJ | 286 nJ | 286 nJ |
| Read Transpose: 64 128kb SRAMs | 2291 nJ | 2291 nJ | 2291 nJ |
| Write: 64 128kb SRAMs | 385 nJ | 385 nJ | 385 nJ |
| Read: 1MB ReRAM | 208 nJ | 208 nJ | 208 nJ |
| Read Transpose: 1MB ReRAM | 208 nJ | 208 nJ | 208 nJ |
| Write: 1MB ReRAM | 676 nJ | 676 nJ | 676 nJ |
| Multiply and Accumulate (1M operations) | 1,500 nJ | 900 nJ | 520 nJ |
| Digital ReRAM Cross Core Communication | 431 nJ | 394 nJ | 370 nJ |
| Digital SRAM Cross Core Communication | 1,065 nJ | 1,051 nJ | 1,042 nJ |
| Totals | | | |
| Analog ReRAM Total | 28 nJ | 2.7 nJ | 1.3 nJ |
| Digital ReRAM Total | 7520 nJ | 5580 nJ | 4340 nJ |
| Digital SRAM Total | 12,010 nJ | 10,150 nJ | 8,970 nJ |

Table IV: Energy Breakdown. The total energy is for a three-step cycle, a VMM, a MVM, and an outer product update.

The level shifter circuitry in Fig. 6 relies on feedback to increase the voltage from the low voltage logic to the higher driver voltage. The feedback-based design is chosen to minimize the transistor count. Circuit simulations calculate that each level shifter and attached driver takes ~200 ps and requires 15 fJ per transition due to the feedback. On average, across the 1024 drivers for 8 bits, this requires 170 pJ during reads. The registers and control logic consumed 35 pJ during reads. During write, the energy is doubled as the drivers must be used for two write cycles.

### C. Voltage Drivers

As with the temporal drivers, the area is dominated by the per column driver. Eight high-voltage (1.8V) transistors are required per-rail that are connected to the voltage-coded inputs (4 transistors per level shifter and 2 drive transistors per array). When including the ground/standby rail, we need $1 + 2^{voltage\_bits - 1}$ rails. We also include some synthesized standard-cell digital logic to choose rails and store the inputs/outputs (Register 2 in Fig. 3). For 8 bits, it has an area of 17 µm² per column and consumes 10 pJ per column. The control logic chooses an appropriate rail if the driver is enabled and the polarity is correct, applies the outputs from the temporal coding to columns, and receives the ADC results and storing them in the included register.

We choose to use $n_{bits,V}$ =4 bits on the voltage driver (3 bits of magnitude + 1 sign bit), as only a few bits are needed for the update. It is important to limit the number of bits here as this is a dominant part of the area cost. Ref. [27] shows that in some cases, 2-bit calculations are sufficient to achieve the same performance accuracy as full double-precision floating point.

The level shifters are identical to those used for the temporal coding. There are $1+2^{voltage\_bits-1}$ level shifters per column. Only the level shifter corresponding to the selected voltage transitions in a given phase use energy giving a total energy cost of 80 pJ regardless of the number of bits.

### D. Integrator

The area of the integrator is estimated from the design in Fig. 8(b). The integrator requires 12 transistors with channel lengths that are 33% longer than the minimum size. The longer channel length increases the area by 19%. There are also 4 minimum sized transistors for the pass gates in Fig. 8(a) giving a total per column area of 6.4 µm². The current input to the integrator will be a maximum of 1 µA.

The size of the integration capacitor depends on the dynamic range required. It would require a 330 fF capacitor to hold the maximum possible charge that is accumulated over 128 ns (7 non-sign bit input) through 1024 devices, with 1 nA at 0.4 V. Fortunately, the dynamic range needed on the outputs is only a few percent of this and a ~10 fF capacitor is required. This is because most of the inputs are zero, or they average to near zero, allowing large values to saturate. Nevertheless, a larger capacitance $C_{load}$ in Fig. 8(a) can be used to minimize the change in the line voltage until the integrator responds and to average the charge over the entire input pulse length. The parasitic capacitance on a column (50 fF) is sufficient for this load and is enough to limit the worst case voltage swing on the column to 10% of the max output voltage if the op-amp does not respond for 2 ns. Circuit simulations indicate that the integrator in Fig. 8(b) has a bandwidth of 5 GHz which is fast enough.

The integrator is run for the same amount of time as the temporal coding drivers. While running it consumes 12 µA of current as verified by circuit simulations. The energy is estimated by taking the maximum input current, multiplying by the maximum voltage (1.8V) and the integration time.

### E. Analog to Digital Converter (ADC)

The ADC consists of a single ramp generator and control logic for the entire array, as well as a comparator for each column (Fig 7). The area of the comparator and associated transistors for the 1024 columns dominates, which is the focus of our calculations. The comparator consists of 13 high voltage transistors, 5 of which are larger than minimum size (Fig. 8(c)). Consequently, the area per comparator is 5.7 µm².

The ramp is switched at 1 level/ns and the total run time is given by the number of ADC levels × 1ns = 256 ns for 8 bits. The energy will be dominated by the 1024 comparators which each consume 20 µA at 1.8V. Current consumption and switching speed were verified by SPICE circuit simulations.

### F. Analog Routing

The array needs to be able to connect and disconnect the drivers and outputs to switch between operations. The drivers in Fig 6(b) can be set to provide a high-Z drive, thereby disconnecting the driver from the row (or column). The positive and negative arrays also must be capable independent drive, which is achieved by driving each array from independent power rails. An array can be deactivated by disconnecting its power rails.

A single integrator is shared between both a row and a column, and four pass gates (2 arrays x 2 pass gates per array) are used to connect the integrator to the desired input. Hence, eight high voltage transistors per column are required.

### G. Digital ReRAM Array

In order to assess the performance of a digital ReRAM-based neural core, the design must be optimized to minimize area and maximize throughput. The density is maximized by considering eight 1024x1024 arrays, providing 8 MB of weight storage. In designing the arrays, throughput must be maximized as all values in an array are read out and written in a single cycle.

The number elements in a row that can be read or written in parallel is limited in a crossbar configuration due to parasitic voltage drops, and by electromigration current limits on a minimum sized wire. We optimize the digital ReRAM memory to operate in the regime where the half-select leakage power does not dominate the read/write energy and can be ignored. To do this, the parasitic voltage drop should not be more than roughly 100 mV [30, 31]. As seen in refs [30] and [31] once the parasitic voltage drops become significant, the write energy increases exponentially, significantly dominating the system power. Using larger wires does not resolve this as both the row and column wires would need to be wider, resulting in the same resistance per memory cell. Dividing 100 mV by the resistance of a row in a 1024x1024 array gives a maximum current of 54 µA. This current sets a limit on the number of rows and columns that can be read and written in parallel. The more devices that are written in parallel, the lower the read current will be and therefore the slower the reads will be. Optimizing such that the time to read or write the entire array is equal results in the binary ReRAM parameters in Table I. The array is read by adding a series resistor equal to $R_{Load} = \sqrt{R_{on} \times R_{off}}$ and measuring whether the voltage across it is above or below a threshold with a sense amp. The load resistance will be very high and consequently will need to be made using a similar process to the ReRAM. The write current per device is 54 µA divided by the number of devices written in parallel. This sets the on-state resistance, which can then be used to find the read current (assuming 0.1V across the ReRAM during a read) and thus the read parallelism (54 µA/ read current). We assume each device needs 10 ns to write and estimate the read time by taking 2.2 ×(RC time) which is estimated as follows [32].

$$\tau_{RC} = \frac{R_{line} \times C_{line}}{2}\left(1 + 2 \times \frac{R_{ReRAM}//R_{Load}}{R_{line}}\right) \quad [5]$$

$R_{line}$ and $C_{line}$ are the resistance and capacitance of the column and $R_{ReRAM}//R_{Load}$ is the parallel combination of the ReRAM and load resistances.

Optimizing to get equal read and write times for the entire array results in 64 bits on a row being written in parallel and 512 bits read in parallel. The write latency is 10 ns and the read latency is 86 ns. The time to write a full array is 164 µs and the time to read an entire array is 176 µs. The read and write energies are found by summing the $CV^2$ and $I \times V \times \tau$ energies. The total read energy is 166 nJ and the write energy is 676 nJ. The energies are dominated by the $CV^2$ energy as the columns need to be charged once per bit or $8 \times 10^6$ times.

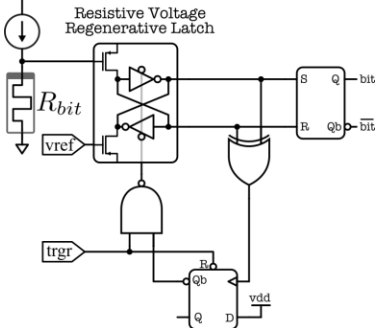

Fig 9: The digital sense amp design is shown. The regenerative latch would flow current after switching and so the logic is used to shut it off after switching. The state is saved in the SR latch.

The array drivers contribute a small amount of energy and delay compared to the array itself. However, the area of the drivers is important. In order to drive the array during read or write, the array can be driven with one of two voltages. Consequently, each row or col will need two level shifters and two drive transistors, or 10 high voltage transistors. We will also need two pass gates to switch from row read output to col. read output that requires 4 high voltage transistors per col. Thus a total of 24 high voltage transistors are needed per col. The read drivers will need a 10:1024 decoder and much smaller 5:32 and 2:4 decoders as well as five low voltage pass gates per row to route the decoder outputs for an area of 200 µm² (based on Verilog synthesis).

In order to read out 256 rows we will need 256 sense amps. The sense amp design is shown in Fig 9 and can be made of low voltage transistors as the output voltage will not exceed 0.8 V. This requires 60 low voltage transistors per sense amp. Thus the total area is 9,500 µm², about twice the array size, and so the array fully fits over its drivers. The sense amp consumes 5 fJ per measurement or 5.2 nJ per array.

### H. Digital SRAM Array

A 1MB cache was synthesized using a cache generator targeting the PDK to give areas, latencies, and energy as shown in Tables II through IV. Due to limitations of the maximum size cache generated by the SRAM generator, we logically combine 64 128 kb generated SRAMS into a single physical array capable of holding the entire matrix. This repetition of address circuitry likely adds a slight area overhead compared to a fully optimized 1MB implementation, but energy and latency should be equal or improved. Each SRAM can read or write 64 bits in 2 ns. Each 128 kb array requires 34 fJ/bit to read, 46 fJ/bit to write, and 12,103 µm². The cross-core communication energy noted in Table III represents energy to transport data from the edge of an instance of the generated cache to the nearby computation units. The reads are pipelined with the multiply and accumulate. It should be noted that digital place and route was not performed, and hence the energy and area for the digital implementation represent a best-case scenario.

Unlike ReRAM crossbars, it is not trivial to implement a dense SRAM that is capable of both row-major and column-major reading. Therefore, to operate on the transpose of the stored matrix, 8X additional reads are required, as the data returned from the SRAM is not otherwise properly aligned with the input vectors being sent to the multiply-accumulate units. (The matrix data is stored in SRAM arranged for row-major access. Non-1D-blocked arrangements were considered, but those result in more "wasted" reads.)

### I. Digital Input Registers

The row inputs are stored in a register for the digital memory arrays. The area is based on 1024x8 standard-cell flip-flops. Because the drivers require bitwise access to the buffers, we cannot utilize a more-conventional register file. The access time is one clock cycle or 1 ns.

### J. Multiply and Accumulate

An 8-bit multiply and accumulate unit was synthesized and the area was multiplied by 256 to give 54,000µm². The multiply is internally rounded to 12 bits of precision and the accumulate is done to 22 bits of precision internally to prevent issues with saturation resulting in skewed results. The result is then rounded/saturated to keep the desired 8 bits. For the 4- and 2-bit input versions, the top 8 bits of the multiply and 18 bits of the accumulate are kept.

The synthesized block operates on a 1 GHz clock. Although each operation requires 2ns to complete, operation is pipelined, with one input every clock cycle., using ~1.46pJ per 8-bit multiply-add operation, including writeback to the buffer.

### K. Cross Core Communication

We add the energy to move each bit in the matrix storage across the core. This is because the memory arrays are designed based on smaller sub arrays and the communication energy to get the data to its destination must be included. For the analog array the drivers are larger than the array and so the extra communication energy for that is needed as well. The communication energy is estimated by finding the $CV^2$ energy to charge a wire equal to the edge length of the core ($\sqrt{area}$) and multiplying by $n_{rows} \times n_{cols} \times 8\ bits$ for digital and ($n_{rows} + n_{cols}$) for analog. We see that these energies can dominate for digital as data movement is very expensive. Optimizing the position of the multiply and accumulate units relative to the memory cache becomes critical.

### L. Discussion

Overall, we see that the analog accelerator offers a significant performance advantage over digital accelerators. Compared to digital ReRAM, the energy, latency and area are 270X, 1040X, and 1.8X better respectively. Compared to an SRAM based accelerator, energy, latency and area are 430X, 34X, and 11X better respectively. The 2-3 orders of magnitude improvements in performance fundamentally come from two analog advantages. Analog accelerators do not have to move every stored memory bit and they get the multiply and accumulates for free. These two costs dominate the digital accelerators, and they are free for analog. These improvements are at the kernel level, a full accelerator architecture must be developed to fully utilize the analog circuit-block advantages.

Furthermore, the low precision 2- and 4-bit results show that analog can gain an additional order of magnitude over digital if algorithms can be designed that only need low precision inputs and outputs (while still keeping ~8-bit weights). This because the biggest cost in analog is the temporal coding, which drops exponentially when the number of input/output bits is reduced. The analog accelerator could be further improved if the area of the high voltage transistors needed to drive the array could be reduced. Currently they are 8X larger than the low voltage transistors. The energy is also limited by the ramp ADC as the comparators burn 20 µA of current for an extended period of time. Lower current comparator designs would also help significantly reduce the energy.

| Component | Vector Matrix Multiply | Matrix Vector Multiply | Outer Product Update |
|---|---|---|---|
| Energy | | | |
| Analog ReRAM Total | 12.8 nJ | 12.8 nJ | 2.2 nJ |
| Digital ReRAM Total | 2140 nJ | 2140 nJ | 3250 nJ |
| Digital SRAM Total | 2850 nJ | 4855 nJ | 4300 nJ |
| Latency | | | |
| Analog ReRAM Total | 0.384 µs | 0.384 µs | 0.512 µs |
| Digital ReRAM Total | 176 µs | 176 µs | 340 µs |
| Digital SRAM Total | 4 µs | 32 µs | 8 µs |

Table V. Overall comparative analysis of energy and latency.

## V. Assessing the Suitability of ReRAM for use as an Analog Neural Accelerator Training Element

One of the major challenges of an analog accelerator is that the algorithm level characteristics are affected by the device. It is not sufficiently accurate to simply represent the device as a low precision number (e.g. 6-bit). The experimental device behavior can be a complicated, stochastic function of current conductance state and direction of change. Hence, a careful procedure of measurement and modeling the analog ReRAM characteristics must be undertaken to accurately predict the algorithm training accuracy, as described in the following.

During training, the weight update is calculated by the algorithm, which in this case is backpropagation. In the simplest case, this value is linearly converted into a specific number of identical voltage pulses which is proportional to the weight change. For more efficient writes the value can also be encoded into pulse lengths and voltages if the change in conductance is well modeled as discussed later. For an ideal device, each identical pulse will cause the same change in conductance. The blue curve in Fig. 10 schematically illustrates a plot of initial device conductance $G_0$ versus conductance change $\Delta G$ for an ideal device. Unfortunately, real resistive switching devices conductance are subject to several effects which are discussed below. Measuring and modeling these effects accurately is paramount to predicting the algorithm training accuracy for the analog accelerator.

### A. Read and Write Nonidealities in ReRAM Weights

ReRAM is experimentally subject to several sources of noise during read and write operations. Read noise has been observed as fluctuations in the current when read at a constant voltage, which is more severe at low currents [33]. Read noise may dominate output precision in an analog system when the weights are precisely set using a feedback system for the write operations. An example analog vector matrix multiply units is the dot product engine, which programs the weights in a crossbar to within 10% of a precise 8-bit value using a closed-loop programming scheme [34].

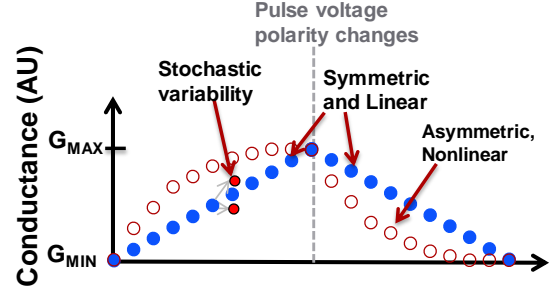

Fig. 10. Illustration of conductance vs. pulse number.

However, read noise in an open-loop training accelerator is not the factor which dominates algorithm accuracy. The effect of read noise is negligible for the case where the magnitude of the fluctuation is less than about 5% of the current [22]. Rather, in the open-loop analog training accelerator, the write nonidealities are the main factor which determines algorithm level accuracy. There are three major write nonidealities which must be considered: i) nonlinearity, ii) asymmetry, and iii) write stochasticity.

Nonlinearity is the dependence of weight change on starting conductance for a given pulse length and amplitude. This is illustrated in the hollow-dotted curve of Fig. 10. In this example, $\Delta G$ is significantly larger at low $G_0$ when the weight is increasing. When the sign of the voltage shifts, the nonlinearity changes nature such that $\Delta G$ is greatest at high $G_0$, as seen in the right half of Fig. 4. This effect is the effect of asymmetry. This means that if multiple positive pulses are applied to a device at high conductance, the conductance will not significantly change, while a single negative pulse will cause a large change undoing the training from multiple previous positive pulses. Nonlinearity and asymmetry have been modeled analytically [35],[22] and experimentally [27], and have a significant effect on algorithm training accuracy.

Write stochasticity is the effect that even in the absence of nonlinearity, $\Delta G$ will fluctuate randomly for a given pulse width and voltage. This effect can be explained using Fig. 10. The blue curve represents the ideal behavior without stochasticity or nonlinearity. In the absence of nonlinearity, each pulse of the blue curve will still result in a random final conductance, whose 3σ bounds are represented by the red dots.

Measuring and modeling these combined effects allows us to predict final algorithm level training accuracy. The main measurement required to measure three effects for a device is a repeated pulse train of fixed voltage and pulse width. This data can then be used to create a lookup table as described below.

### B. TaO$_x$ ReRAM Experimental Details

Sandia's standard semi-production Ta/TaO$_x$ bipolar ReRAM cells described in Section II were used to assess the analog accuracy and voltage-pulse dependence of ReRAM. These behave similar to other oxide-based bipolar ReRAM operated in the same conductance range, and hence are likely representative. However, cells do not represent the record endurance or maximum resistance reported, so those characteristics are assessed with respect to the literature.

Pulsed measurements were made using an Agilent B1500 with a B1530 fast waveform generation unit. Device were measured on a low noise Cascade RF probe station using ground-signal probes. Pulse rise times as fast as 10 ns are possible with this setup. Current measurements were made with a Keysight CX3300 current analyzer.

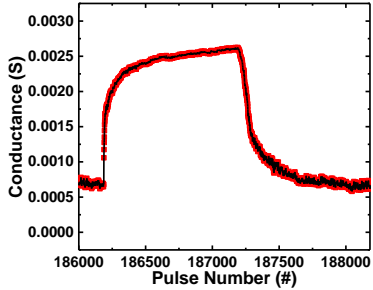

Fig. 11. Conductance versus pulse number for analog cycling of Sandia TaOx ReRAM cell.

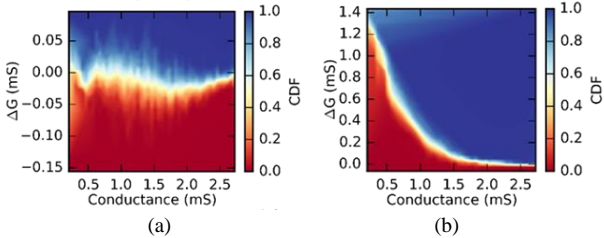

(a)  (b)

Fig. 12. ΔG versus $G_0$ for (a) RESET and (b) SET processes for analog cycling of Sandia TaO$_x$ ReRAM cell.

An automated measurement routine consisting of 1000 pulses with positive polarity on $V_{TE}$ (shown in Fig. 1) followed by 1000 pulses of negative polarity is used, replicating how the ReRAM weight is operated in an analog accelerator. The data resulting from a single cycle is plotted in Fig. 11. In order to gain a sufficiently averaged dataset, this entire process is repeated 1k to 10k cycles, resulting in 1M to 10M total pulses. If wearout effects are to be modeled as well, it is necessary to cycle the device until a significant narrowing of the $G_{MAX}/G_{MIN}$ window occurs.

### C. Conductance Change Characteristic Dataset

Nonlinearity, asymmetry, and stochasticity effects described above all can be observed in the typical plot of G versus pulse number (G-pulse), given in Fig. 11 for TaO$_x$ ReRAM. In order to model these effects on the training of the algorithm, statistical data is extracted from the repeated pulsing between $G_{MIN}$ and $G_{MAX}$ following the methodology first presented in Burr *et al* [27, 36]. The complete G-pulse data for a single pulse amplitude and width is sorted into bins which represent the conductance immediately before a pulse was applied. For each of these bins, the distribution of ΔG resulting from the pulse is mapped as a probability distribution. This produces a heat-map of ΔG versus $G_0$ format, as illustrated in Fig 12. Each plot represents a single pulse amplitude and length; in this case the SET plot is $V_{TE}$=+1V and RESET $V_{TE}$=-2.5.

### D. Conductance Change versus Voltage Behavior

The conductance versus pulse must follow a predictable behavior with voltage for the write scheme used in Fig. 3(c) to be viable. This has been examined by measuring ΔG versus pulse voltage for a range of SET and RESET voltages, as plotted in Fig. 13. This shows a predictable exponential dependence of change in conductance on voltage. In particular, the relationship can be described by:

$$\Delta G(V) = \begin{cases} e^{d_1 \times (V - V_{min,p})} - 1 & V > V_{min,p} \\ e^{d_2 \times (V_{min,n} - V)} - 1 & V < V_{min,n} \\ 0 & V_{min,n} < V < V_{min,p} \end{cases} \quad [6]$$

where $d_1$ and $d_2$ are device dependent properties, $V_{min,p}$ and $V_{min,n}$ are the minimum positive and negative voltages required to change the device state. Examples of $V_{min,p}$ and $V_{min,n}$ are indicated in Fig. 13 for the TaOx ReRAM. This indicates the voltage encoding scheme above is viable for this device.

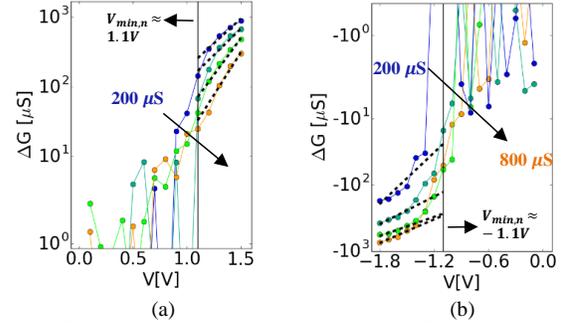

(a)  (b)

Fig. 13. Dependence of ΔG on pulse voltage for a fixed pulse width for (a) SET (positive) (b) RESET (negative) voltage pulses.

### E. ReRAM Endurance and Wearout Effects

Wearout is an important consideration for training acceleration. Each write operation described above is able to "nudge" the cell. It is estimated that the device will be trained at a rate on the order of 100 kHz. During a single training cycle, the device may experience as many as 256 voltage pulses (for the 8-bit scheme). Hence, continuous operation for one year requires an endurance of ~$8\times10^{14}$ single pulses. However, it is unlikely that a device would experience the greatest number of pulses (i.e. 256) in a cycle or be even be updated each training cycle. If we assume on average that the device experiences an average of 128 pulses 10% of cycles, the required number of single pulses is ~$4\times10^{13}$.

TaOx-based ReRAM cells in the literature have been shown to reach $10^{12}$ cycles when operated as a memory [19]. In this case, each cycle represents a full swing from $G_{MIN}$ to $G_{MAX}$ and back to $G_{MIN}$, after which it is repeated. In our training scheme, each of these cycles is considered two "nudges". Therefore, it can be considered that devices from the literature have achieved the equivalent of $2\times10^{12}$ updates.

The energy involved in a device nudge is significantly less than the full change from the LRS to HRS memory state, and hence device nudge endurance might be significantly greater than full endurance. A test of endurance and analog cycling to failure using statistically relevant data is needed to prove this.

## F. ReRAM Write Physical Current Constraints

In order to update the entire crossbar in parallel, maximize energy efficiency and minimize latency, all devices must be able to write simultaneously. This leads to the requirement that the maximum switching current does not exceed the electromigration current limits of the wire. If the 14/16 nm node is considered, the metal-1 lines are 32 nm, and have a current limit of $I_{limit} \approx 33$ µA. If $N$ devices on a column are switched simultaneously, with the maximum individual "nudge" write current as $I_{nudge}$, then the column line will experience a current $I_{column}= I_{nudge} \times N$. Hence, electromigration limits require that $I_{column} N < I_{limit}$. For the case of a 1024x1024 array, N=1024, and this gives a maximum allowable $I_{nudge} \approx 32$ nA, corresponding to an approximate minimum resistance of $R_{ON} \approx 31$ MΩ. As discussed above, we have limited $I_{nudge}$ to 10 nA to limit the parasitic voltage drop to <20mV. In the case that a greater voltage drop was acceptable this is the absolute physical limit.

## VI. EVALUATION OF ReRAM CROSSBAR ACCURACY IN NEURAL NETWORK TRAINING

### A. Accuracy of MNIST Training

Using the dataset plotted in Fig. 12, it is possible to simulate the accuracy of a training algorithm when ReRAM weights are used. We examined this using our open-source CrossSim code[5] for the case of using backpropagation to train the MNIST dataset [6]. This uses a three layer network configured as 784x300x34. During training in backpropagation, after each weight update is calculated, a CrossSim module is called which adds the combined effect of the nonidealities described above.

A typical plot of accuracy versus epoch for the Sandia TaO$_x$ ReRAM device is plotted in Fig. 14, which is compared with the accuracy possible using single precision floating point values for training (labeled the *numeric* curve). The maximum accuracy possible with the device is about 77%, whereas training the network numerically is about 98%. It is possible, by optimizing the pulsing length and voltage to obtain accuracies up to ~85%. This level of accuracy loss is not acceptable for most applications.

It is useful to examine the relative sources of degradation. The red curve has all degradation mechanisms. The *no-noise* curve is without stochasticity, such that the change in conductance follows a deterministic nonlinear path. The *linearized* curve removes the state dependence from the TaO$_x$ ReRAM pulse curve by assuming each device is serially written based on its state. This nonlinear dependence of conductance change on starting state is clearly the greatest degrader of accuracy.

### B. Accuracy Improvements

Several methods have been proposed by which the accuracy can be improved. One method is to use multiple ReRAM cells for each weight. We have developed an efficient version of this technique called *periodic carry*, which uses multiple synapses where each synapse is used to represent increasing significance in a place value based number (i.e. base 10) representing the weight while maintaining the benefit of a parallel update [37].

Using period carry allows an analog TaO$_x$ ReRAM to reach within 1% of numerical accuracy (Fig. 15).

In principal, any two-terminal electronic device can serve as a weight element if it meets these analog requirements. We have evaluated Cu/SiO$_2$ CBRAM devices and found accuracy similar to TaO$_x$. The emerging electrochemical Nonvolatile Redox Transistors (NVRT) have been shown to significantly improve accuracy to near numerical accuracy [16]. Hence, this is a promising direction if speed and endurance can be proven.

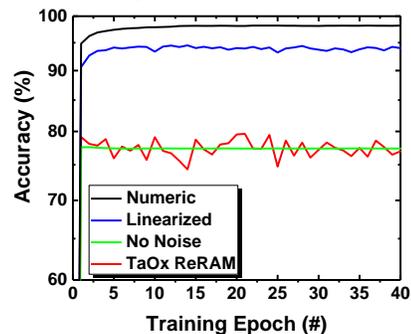

Fig. 14. Accuracy versus training epoch for TaO$_x$ ReRAM versus numeric with all models, and different error models activated.

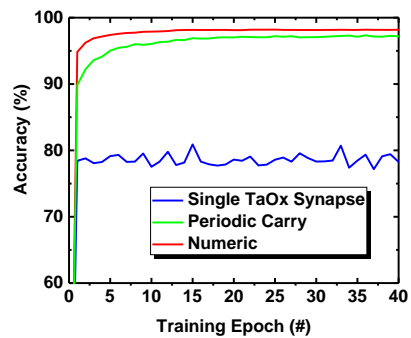

Fig. 15. Accuracy versus training epoch compared for a floating point (red curve), single TaO$_x$ ReRAM (blue curve), and 3-device TaO$_x$ ReRAM with period carry implemented (green curve) [36].

## VII. DISCUSSION AND FUTURE CHALLENGES

This analysis of energy, latency, area, and accuracy shows that an analog neural training accelerator has significant potential advantages compared to an SRAM/CMOS-only and ReRAM memory-only accelerator. Compared to digital ReRAM, the energy, latency and area are 270X, 540X, and 1.8X better respectively. Compared to an SRAM based accelerator, the energy, latency and area are 430X, 34X, and 11X better respectively. Furthermore, an analog multiply-accumulate requires ~11 fJ, meeting the target specified above.

However, significant unsolved challenges remain; those of greatest significance are:

1. **ReRAM and other resistance change devices do not yet meet the combined analog requirements**. Ultra-low current devices have been presented in the literature [20] but it is not clear that these can meet endurance requirements. Furthermore, analog behavior for ultra-low current devices has not been reported.
2. **ReRAM has not yet met the endurance requirement to sustain training operation for the necessary 1-5 years**. Devices have been presented with $10^{12}$ memory cycles [19], whereas $>10^{13}$ is required. Smaller analog nudge updates may have longer endurance, but this needs to be proven.

3. **The voltage of a resistive switching device in series with an inline select device necessitates high voltage transistors.** Reducing the combined voltage to about 800 mV will enable the use of standard 14/16nm FinFETs and greatly reduce the area.

4. **Algorithm and architecture must be redesigned to realize the maximum efficiency gain from an analog accelerator.** The biggest cost in analog ReRAM is the temporal coding for higher precision inputs/outputs. Algorithms that can operate with 2-4 bit inputs/outputs and 8 bit weights can easily realize an additional order of magnitude improvement.

## VIII. Conclusion

Energy efficiency, accuracy, latency, area of a ReRAM-based analog training accelerator block have been analyzed and compared to digital SRAM and ReRAM implementations. The analog ReRAM accelerator provides an energy gain over a CMOS-only accelerator of about 430x, and about 270x better than with digital ReRAM, due largely to the reduced data movement and free analog vector-matrix operations. Significant improvements in analog device properties are still required to realize these gains.